\documentclass{emulateapj}

\usepackage{amsmath}
\usepackage{graphicx}
\usepackage{float}
\usepackage{amsmath}
\usepackage{epsfig,floatflt}
\usepackage{subfigure}
\usepackage{apjfonts}

\def \approxgt{\,\raise2pt \hbox{$>$}\kern-8pt\lower2.pt\hbox{$\sim$}\,}
\def \approxlt{\,\raise2pt \hbox{$<$}\kern-8pt\lower2.pt\hbox{$\sim$}\,}
\def \th{\thinspace}
\def \ngth{\negthinspace}
\def \ngth2{\negthinspace\negthinspace}
\def \ni{\noindent}
\def \Teff{{$T_{\rm {ef\!f}} $}}

\def\Lo{{$L_\odot $}}
\def\Mo{{$M_\odot $}}
\def\Ro{{$R_\odot $}}
\def \at{{\rm\char'100}}

\def \eg{{{\it e.g.},\ }}
\def \etal{{\it et al.\ }}

\def \cf{{\it cf.\ }}

\def \ie{{{\it i.e.},\ }}

\def\Log{{\mathrm {Log}}}

\long\def\jumpover#1{{}}

\shorttitle{Ultra-Low Amplitude Cepheids}
\shortauthors{Buchler \etal}

\begin{document}



\title{Ultra-low Amplitude Cepheids in the Large Magellanic Cloud}
\author{J. Robert Buchler\altaffilmark{1,2}, 
Peter R. Wood\altaffilmark{2},
Stefan Keller\altaffilmark{2}
\& Igor Soszy{\'n}ski\altaffilmark{3,4}}
\altaffiltext{1}{\ni Physics Department, University of Florida, 
Gainesville, FL 32611, USA; \\ buchler\at phys.ufl.edu}
\altaffiltext{2}{\ni Research School 
of Astronomy \& Astrophysics, Australian National University, Canberra, 
AUSTRALIA,
wood\at mso.anu.edu.au}
\altaffiltext{3}{Warsaw University Observatory, Al. Ujazdowskie 4, 00-478
         Warszawa, POLAND,
soszynsk\at astrouw.edu.pl}
\altaffiltext{4}{Universidad de Concepcion, Departamento de Fisica, Casilla
         160--C, Concepcion, CHILE}

\begin{abstract} The MACHO variables of LMC Field 77 that lie in the vicinity
of the Cepheid instability strip are reexamined.  Among the 144 variables that
we identify as Cepheids we find 14 that have Fourier amplitudes $<$ 0.05\th mag
in the MACHO red band, of which 7 have an amplitude $<0.006\th$mag : we dub the
latter group of stars ultra-low amplitude (ULA) Cepheids.  The variability of
these objects is verified by a comparison of the MACHO red with the MACHO blue
lightcurves and with those of the corresponding OGLE LMC stars.  The occurrence
of ULA Cepheids is in agreement with theory.  We have also discovered 2 low
amplitude variables whose periods are about a factor of 5--6 smaller than those
of F Cepheids of equal apparent magnitude.  We suggest that these objects are
Cepheids undergoing pulsations in a surface mode and that they belong to a
novel class of Strange Cepheids (or Surface Mode Cepheids) whose existence was
predicted by Buchler \etal (1997).  \end{abstract}





\section{Introduction}

The MACHO, EROS and OGLE data of the Large Magellan Cloud (LMC) have already
been searched for variables, and in particular classical Cepheid variables (\eg
Beaulieu \etal (1995), Welch \etal (1995), Udalski \etal (1999), Kanbur \etal
(2003)).  The lowest reported amplitudes in these analyses have been around
0.01 mag.  Our purpose for redoing such an analysis is to detect as many
Cepheids as possible, with particular emphasis on those with very small
pulsation amplitudes.  Such low amplitude stars are expected from an
evolutionary point of view, either just ramping up their amplitudes after
entering the instability strip (IS), or decaying after exiting the strip.  More
details will be given in the discussion.  Our analysis finds that the MACHO and
OGLE data are accurate enough to go down to several milli-magnitude pulsations.

We have examined the 636 MACHO stars of Field 77 in a parallelogram in the HR
diagram defined by $14 < V < 16$ and 
$17.64 < V +16.39\th (V-R_{\rm c})< 24.03$
This region was chosen by visual inspection to include the instability
strip and colors 0.15 blueward and redward.  It contains a mixture of
non-oscillatory giants of spectral type F, and variable stars such as Cepheids,
W~Vir stars, and ellipsoidal variables (binaries).  This region was converted
from Johnson $V$ and Cousins $R_{\rm c}$ into MACHO blue magnitude (M$_{\rm
B}$) and red magnitude (M$_{\rm R}$) using the transforms given in Alcock \etal
(1999).

Fourier analysis is known to be very good at detecting periodicity in datasets
even in the presence of large noise.  We have performed a Fourier analysis of
the MACHO $M_R$ and $M_B$ datasets of the 636 objects with MUFRAN
(multi-frequency analysis, Koll\'ath 1990) in the frequency range 0.02 -- 0.98
d$^{-1}$.
  
We first reduced the set of objects to those in which there are coincidences
among the 8 largest Fourier peaks in MACHO $M_R$ and $M_B$.  For the usual,
large amplitude Cepheids the peaks are extremely sharp and these Cepheids are
thus readily identified.  Interestingly, there are a number of objects for
which the peaks are not very pronounced, but nevertheless there are
coincidences among the highest Fourier peaks. Each of these cases has had to be
examined individually to ascertain that the detected variability is not
spurious.  Independently, we have used the Phase Dispersion Minimization
routine PDM in IRAF (Stellingwerf 1978) to confirm the detected common MUFRAN
frequency.

How can we be sure that the detected variability is real?  One of the tests
already mentioned is to compare the amplitude spectra of the red and the blue
MACHO lightcurves which are to a large degree independent of each other.
However, since both datasets were obtained in common observing conditions and
at the same sampling points in time, they could potentially have a common
spurious periodicity.  Therefore, as a completely independent test, we have
performed a Fourier analysis of the $I$ lightcurves obtained from OGLE-II and
OGLE-III observations.

One possible source of a spurious common signal in both the MACHO and OGLE data
could be a nearby large-amplitude Cepheid.  In order to eliminate this
possibility, we have checked the online MACHO variable star catalog and the
OGLE-II catalog of variable stars in the LMC (Zebrun \etal 2001) for any
variable within 8$^{\prime\prime}$ of the low amplitude Cepheids.  In no case
did we find a nearby Cepheid of similar period.

To summarize then, an object with marginal periodicity had to show a prominent
Fourier peak in OGLE $I$ as well as in MACHO red amplitude spectra and, in
addition, it had to have a peak in MACHO blue at the same frequency.  One final
requirement was that the phase, as well as the period, of all three lightcurves
was the same.

\section{Results}
%
\begin{figure*}
\begin{center}
 \vbox{\epsfxsize=10.8cm\epsfbox{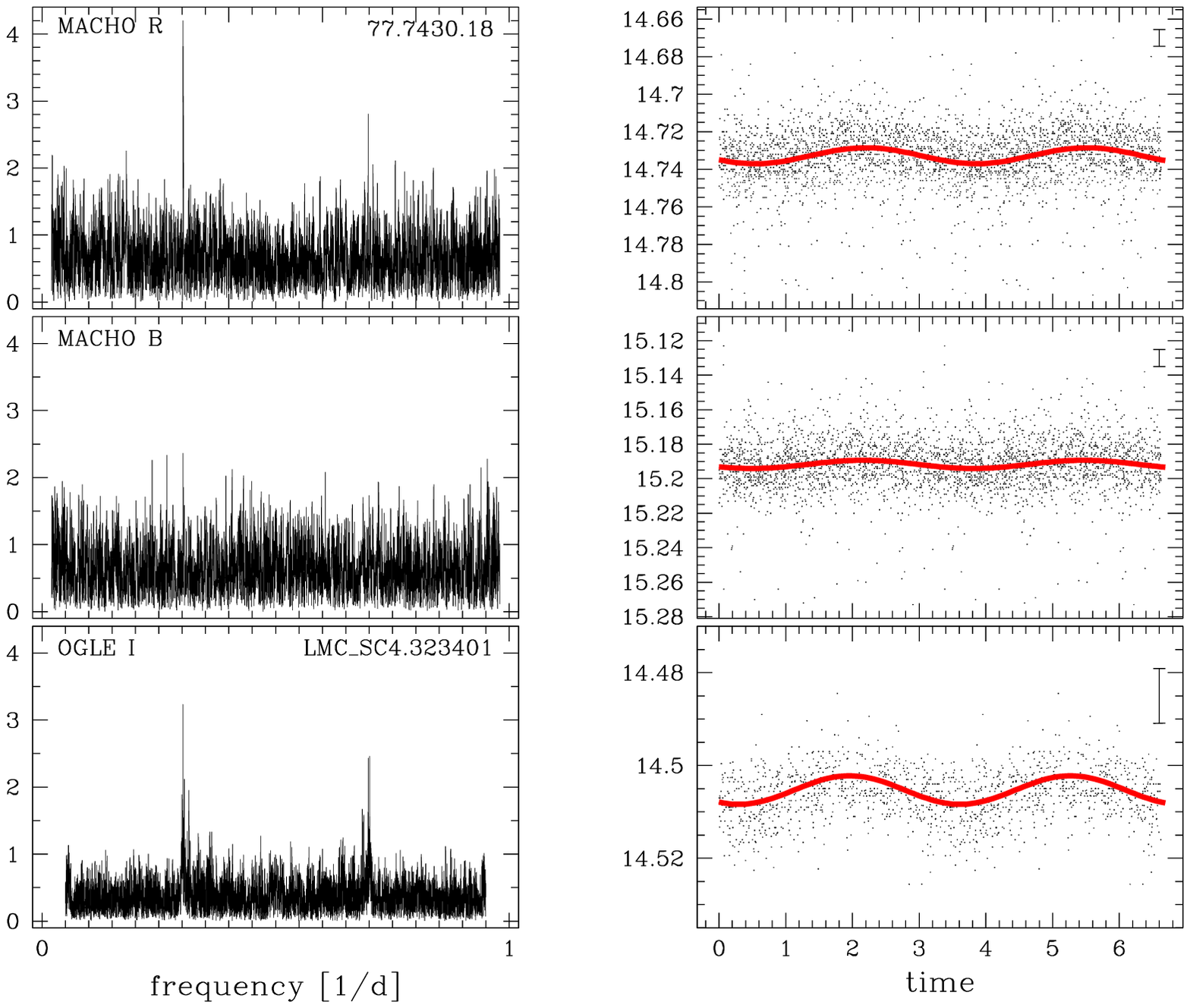}}
\vspace{-8pt}
\end{center}
\noindent{\small Fig.~1: 
Comparison of MACHO Red and Blue (star 77.7430.18) and
OGLE $I$ (SC4 323401) amplitude Fourier spectra [millimag] 
and folded lightcurves.
The best-fit common period is 3.3121\th d.
A $\pm \sigma$
errorbar is shown with the lightcurves
(the $\sigma$ value is an average of the individual
errors given with the MACHO and OGLE lightcurve data).
For this star, the MACHO error estimate appears to be about a factor of 2
smaller
than the true error.
\vspace{-5mm}}
\end{figure*}
In Fig.~1 we present an example of the comparison of the $M_{\rm R}$ and
$M_{\rm B}$ data of the MACHO star 77.7430.18 and $I$ data of the corresponding
OGLE star SC4 323401.  The Fourier amplitude spectra of both the MACHO red and
OGLE data show a very sharp peak near $f_{\rm o}$ =\th 0.30194\th d$^{-1}$
(3.3121\th d). The MACHO blue data appear to be much noisier, and by themselves
would be rejected as just noise, but there is a peak at the same frequency.
Note that the second sharp peak in the Fourier spectra, located at $1-f_{\rm
o}$, is the result of aliasing because of a large, but very sharp window peak
at 1\th d.

The right panels show the data folded with the common period of 3.3121\th d.
This common period has been determined by concatenating the MACHO red and the
OGLE $I$ data and adjusting the zero-point of the OGLE data (on a yearly basis
because of the sizeable yearly shifts in the OGLE data.)  A comparison of the
panels shows that the 3 datasets are in phase and that the
variability is real.  This is further confirmed by least-squares fitting of
single frequency sinusoids to the data: the fits are displayed as thick solid
lines.  Thus, in all respects, this star appears as a very clear case of an
ultra low amplitude (ULA) Cepheid.


\begin{figure*}
\begin{center}
\vbox{\epsfxsize=10.8cm\epsfbox{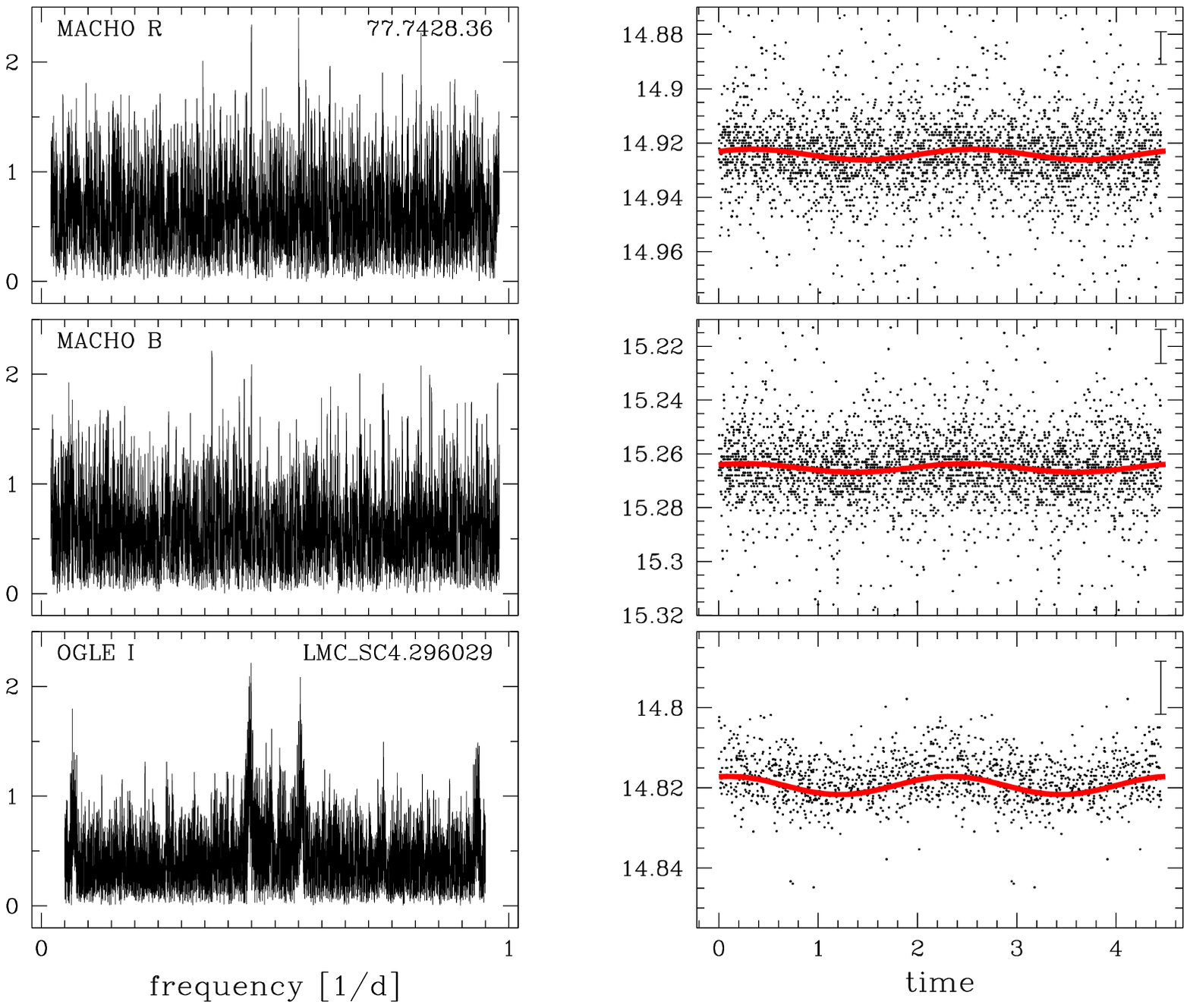}}
\end{center}
\vspace{-9pt}
\noindent{\small Fig.~2: 
Comparison of MACHO Red and Blue (star 77.7428.36) and 
OGLE $I$ (SC4 296029) amplitude Fourier spectra [millimag] 
and folded lightcurves. The best-fit common period is 2.2256\th d.
The errorbars are as in Fig.~1.
\vspace{-5mm}}
\end{figure*}

Fig.~2 gives a second example of the comparison of the $M_{\rm R}$ and $M_{\rm
B}$ data of the MACHO star 77.7428.36 and of the corresponding OGLE star SC4
296029.  The Fourier amplitude spectra of both the MACHO red and OGLE $I$ data
show a common, relatively sharp peak at $f_{\rm o}$ =\th 0.4495 d$^{-1}$ (2.224
d).  The MACHO blue amplitude spectra appear noisier, although the common peak
is the second largest in the blue spectrum.  The diagrams on the right show the
folded lightcurves together with the single frequency (sinusoidal) fit.  Note
that each of the 3 datasets on its own merit would be considered weak or
marginal, but the occurrence of a common peak in all three, and the phase
correlation between the oscillatory parts provides strong evidence that the
periodic variability is real.

Following the above procedures we have identified the Cepheids in MACHO Field
77.  Our 10 lowest amplitude Cepheids are displayed in the Table.  The
identification of periodic variability is solid for all objects.
As far as the quoted amplitudes are concerned, they should be considered upper
limits because the observational noise adds power to the peak.  We estimate
that the actual amplitudes could be 10--20\% lower in the noisiest cases.

\begin{table}
\caption{ten lowest amplitude cepheids in macho lmc field 77
  and corresponding ogle stars}
\vspace{-10pt}
\begin{center}
\begin{tabular}{l  l @{\hspace{2mm}} l @{\hspace{2mm}} r@{\hspace{2mm}}
r@{\hspace{2mm}} r@{\hspace{2mm}} l @{\hspace{2mm}} l}
    \hline
    \hline
    \noalign{\smallskip}
MACHO & &OGLE & A$_R^{\thinspace \dagger}$~ & A$_I^{\thinspace \dagger}$~ 
 & P\th [d]~~   & $M_{\rm B}$ &\quad $M_{\rm R}$ \\
    \noalign{\smallskip}
    \hline
    \noalign{\smallskip}
77.7428.36  & ULA & SC4 296029  & 2.3 & 2.2 & 2.2256  & 15.431 & 15.180 \\
77.7307.21  & ULA & SC4 62503   & 3.3 & 2.7 & 9.7520  & 15.151 & 14.742 \\
77.8032.23  & ULA & SC3 393050  & 3.8 & 2.2 &10.3759  & 15.285 & 14.898 \\
77.8157.16  & ULA & SC2 198696  & 4.0 & 1.8 & 4.8354  & 14.948 & 14.647 \\
77.7430.18  & ULA & SC4 323401  & 4.2 & 3.2 & 3.3121  & 15.334 & 15.012 \\
77.7789.25  & ULA & SC3 153959  & 4.4 & 3.4 & 9.9849  & 15.063 & 14.678 \\
77.7668.981 & ULA & SC3 35239   & 5.4 & 4.1 & 4.3383  & 15.043 & 14.755 \\
77.7306.43  &     & SC4 176301  & 10.9 & 3.2 & 6.7622  & 15.095 & 14.899 \\
77.6940.14  & S   & SC5 124597  & 25.3 & 13.1 & 1.3199  & 15.055 & 14.637 \\
77.7548.11  & S   & SC4 295930  & 31.2 & 25.0 & 6.0370  & 14.139 & 13.548 \\
    \noalign{\smallskip}
    \hline
    \noalign{\vspace{-2.5mm}}
\noalign{\flushleft $^\dagger$ Fourier amplitudes [in millimag] 
for MACHO $M_{\rm R}$ and OGLE $I$ \\
\vspace{-8pt} \flushleft ULA: \th \th Ultra Low Amplitude Cepheid;\quad
 S:  \th \th  Strange Cepheid candidate}
    \noalign{\smallskip}
    \hline
    \hline
\end{tabular}
\end{center}
\label{table1}
\vspace{-12mm}
\end{table}
In Fig.~3 we present our results in an amplitude histogram.  Among the 144
variables that we have identified as Cepheids we find 14 that have amplitudes
$<0.05$\th mag in $M_{\rm R}$, of which 7 have an amplitude $< 0.006$ mag: we
call the latter group of stars ULA Cepheids.  It is possible that out of
caution we have discarded some additional ULA Cepheids because the signal to
noise was too small.

Fig.~4a exhibits the period--luminosity (PL) relation.  To guide the eye we
have added the slanted line, which is approximately parallel to the fundamental
mode blue edge.  It is defined by $W$ = $\alpha \Log\th P$ + $W_{\rm o}$, with
$\alpha\th$ = --3.3, $W_{\rm o}\th$ = --16.01, where $W = M_{\rm R} - 4 \th
(M_{\rm B}-M_{\rm R})$ is a reddening corrected magnitude (e.g. Alcock \etal
1995).


\begin{figure}
\begin{center}
\vbox{\epsfxsize=7.cm\epsfbox{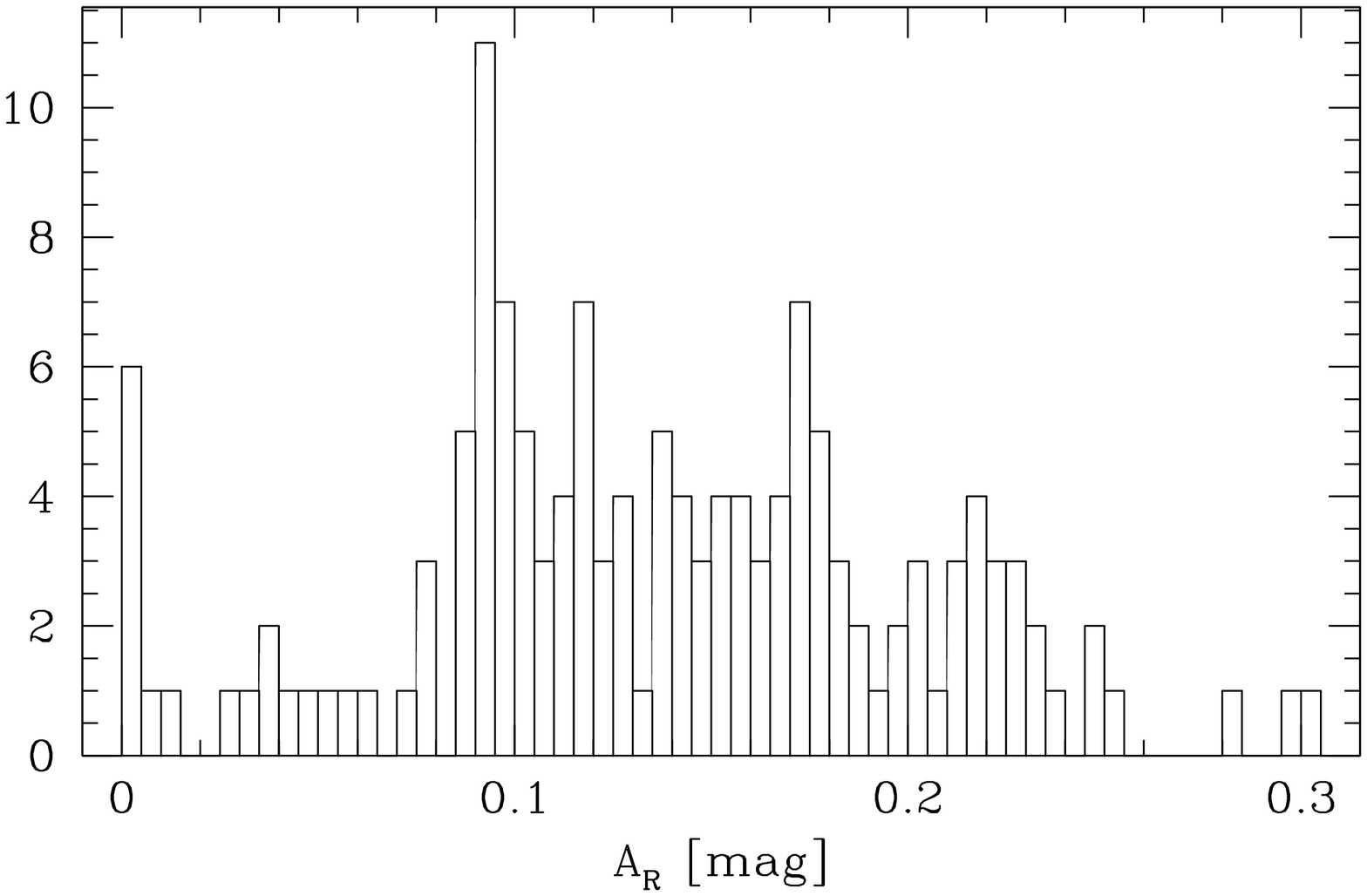}}
\end{center}
\vspace{-8pt}
{\small Fig.~3: Amplitude histogram.}
\vspace{-5mm}
\end{figure}

Fig.~4b gives the period--amplitude relation for all the Cepheids. The crosses
represent the Cepheids that lie below the slanted line of Fig.~4a.


\begin{figure*}
\begin{center}
\vbox{\epsfxsize=7.9cm\epsfbox{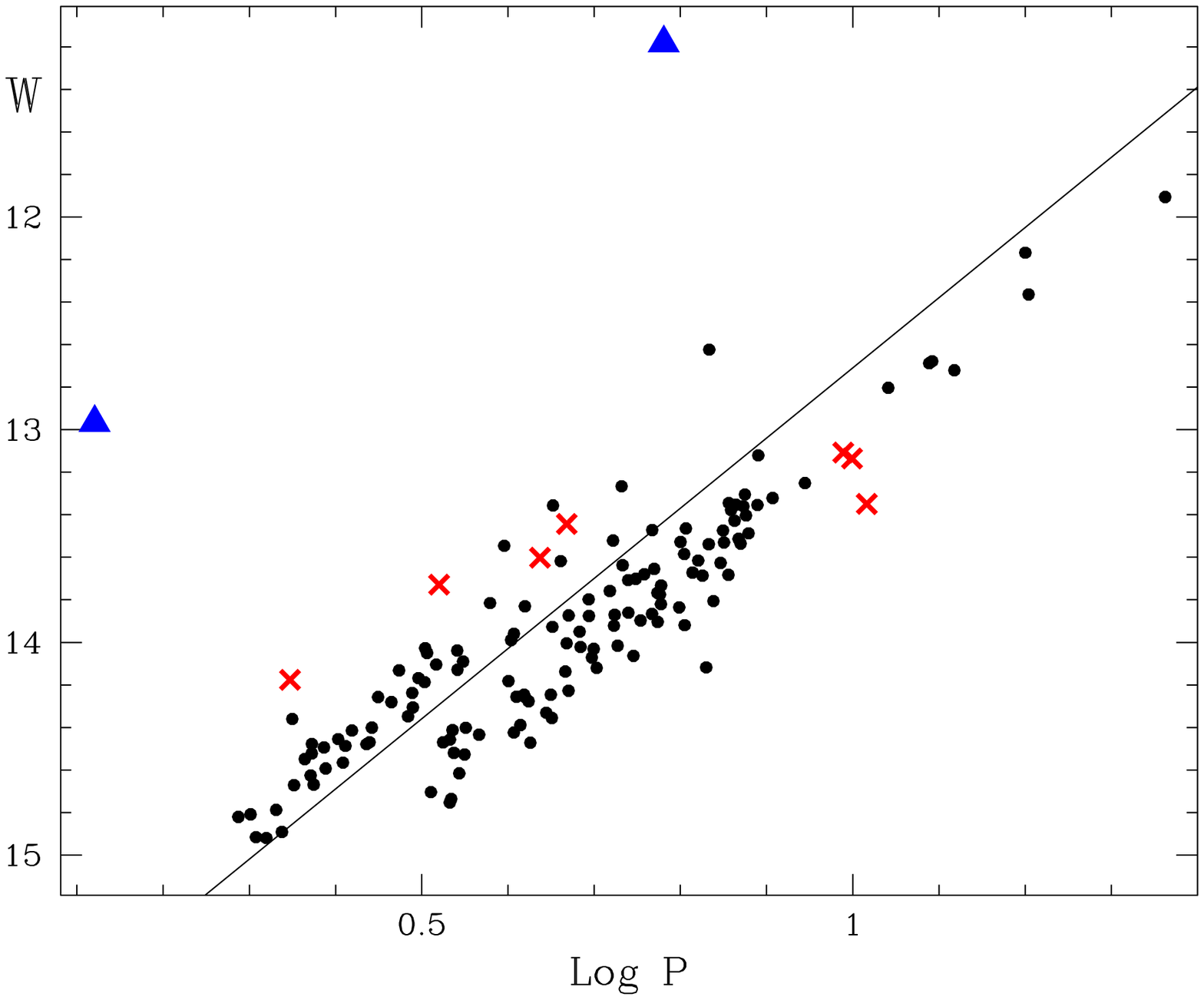}\hspace{5mm}\epsfxsize=7.9cm\epsfbox{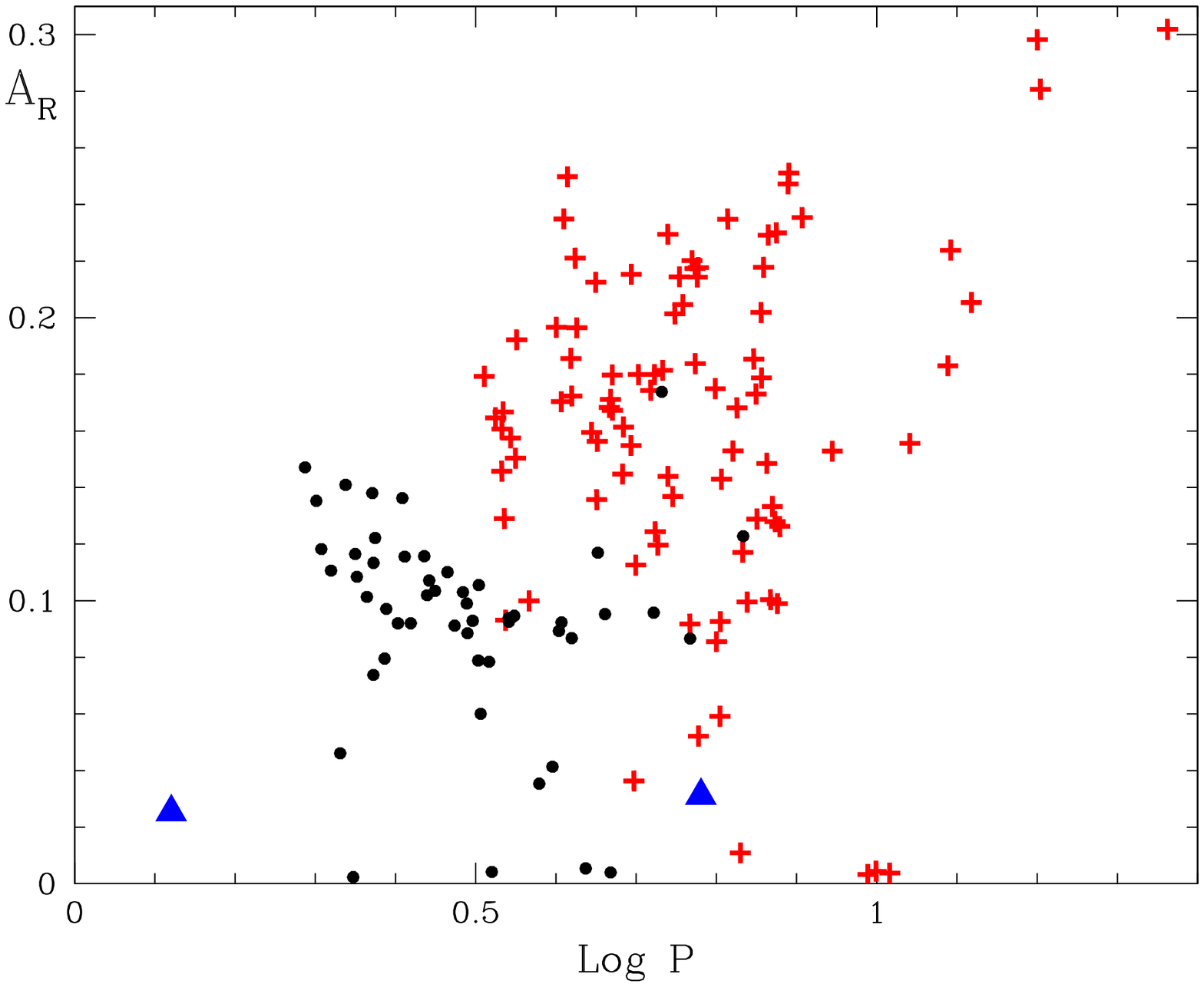}}
\end{center}
\vspace{-8pt}
\noindent{\small Fig.~4: 
(a): Left: Period-Luminosity diagram.  The 7 ULA Cepheids, 
with an amplitude
less than 0.006, are indicated by X's.  The slanted line
approximately separates fundamental from
first overtone pulsators.
The egregious triangles will be
identified as Strange Cepheids.\\
(b) Right:Period- Amplitude diagram. 
The stars that lie below the slanted line of Fig.~4a 
are shown by crosses.
}
\end{figure*}

On the grounds that a time related to pulsation, \eg the amplitude growth time
$\tau_{gr}$, is much shorter than the stellar evolution time $\tau_{evol}$, one
might expect a star to always achieve its full limit cycle pulsation amplitude
$A_{LC}$, \th \ie the amplitude that a standard hydrodynamics code would
compute.  If that were indeed the case then the detection of any Cepheids with
very small amplitudes upon entering or exiting the IS would be very unlikely,
because at the boundaries of the IS the limit cycle amplitude sets in,
respectively decays, with a vertical slope ($dA_{LC}/dt = \infty$;\th \cf
Fig.~5.)  An amplitude histogram would therefore be expected to be devoid of
small amplitude Cepheids.

However, we have identified a surprisingly large number ($\approx 5\% $) of ULA
Cepheids ($<0.006\th$mag).  These stars seem to fall preferentially near the
edges of the Cepheid IS in a P--L diagram.  (Ideally, a color--magnitude plot
should also show this feature, but it has too much scatter because color
is uncertain by at least $\pm$0.05 mag due to reddening variations
(Keller \& Wood 2002) while the whole width of the IS is less than 0.2 mag.)
It turns out that a recent theoretical development predicted the existence of
such ULA Cepheids (Buchler \& Koll\'ath 2002), and provided the incentive for
this search.

At the edges of the IS the amplitude growth rate of a star actually vanishes,
so that $\tau_{gr}$ can be much {\it longer} than $\tau_{evol}$ in the
immediate vicinity of the IS.  As a consequence, as the star enters the IS, the
amplitude does not immediately achieve its full pulsation amplitude.  Rather it
stays at a very low amplitude for a time of the order of
$\sqrt{\tau_{evol}\th\times\th\tau_{gr}}$, but then rapidly grows to full
amplitude.  Therefore very few Cepheids should be detectable during that rapid
amplitude growth, and one would expect to find a gap in the amplitude
distribution between ULA Cepheids and the (usual) full limit cycle amplitudes.
On their way out of the IS, for the same reason the amplitude should not decay
to zero right at the edge of the IS, but the stars should linger in a state of
small amplitude. This behavior is sketched in Fig.~5, on the left for a star
entering the IS, and on the right for a star leaving the IS.  Stellar evolution
calculations further show that Cepheids cross the IS in both directions.  ULA
Cepheids should therefore be found in the vicinity of both the blue and the red
edges of the IS.  (Note that a similar behavior is predicted for RR~Lyrae.)

The existence of ULA Cepheids in the vicinities of the blue and red edges is
thus fully compatible with theoretical predictions (when they are interpreted
as radial pulsators).  A more quantitative comparison of the Cepheid amplitude
distribution should be made with the construction of a synthetic HR diagram
that would combine stellar evolutionary tracks with the results of nonlinear
hydrodynamic simulations along the lines of Szab\'o, Koll\'ath \& Buchler
(2004).

A second theoretical development is the prediction by Buchler \etal (1997) of
the self-excitation of radial overtone modes in Cepheids and RR~Lyrae that are
predominantly surface modes.  They dubbed these modes 'strange modes' because
of their similarity to the strange modes found by Wood (1976) in high
luminosity stars (see also Saio, Wheeler \& Cox, 1984).  This brings us to a
discussion of the two egregious stars, namely 77.6940.14 and 77.7548.11.  They
appear as numbers 9 and 10 in the Table and are indicated by triangles in
Figs.~4a--b.
%
%
\begin{figure}
\begin{center}
\vbox{\epsfxsize=8.6cm\epsfbox{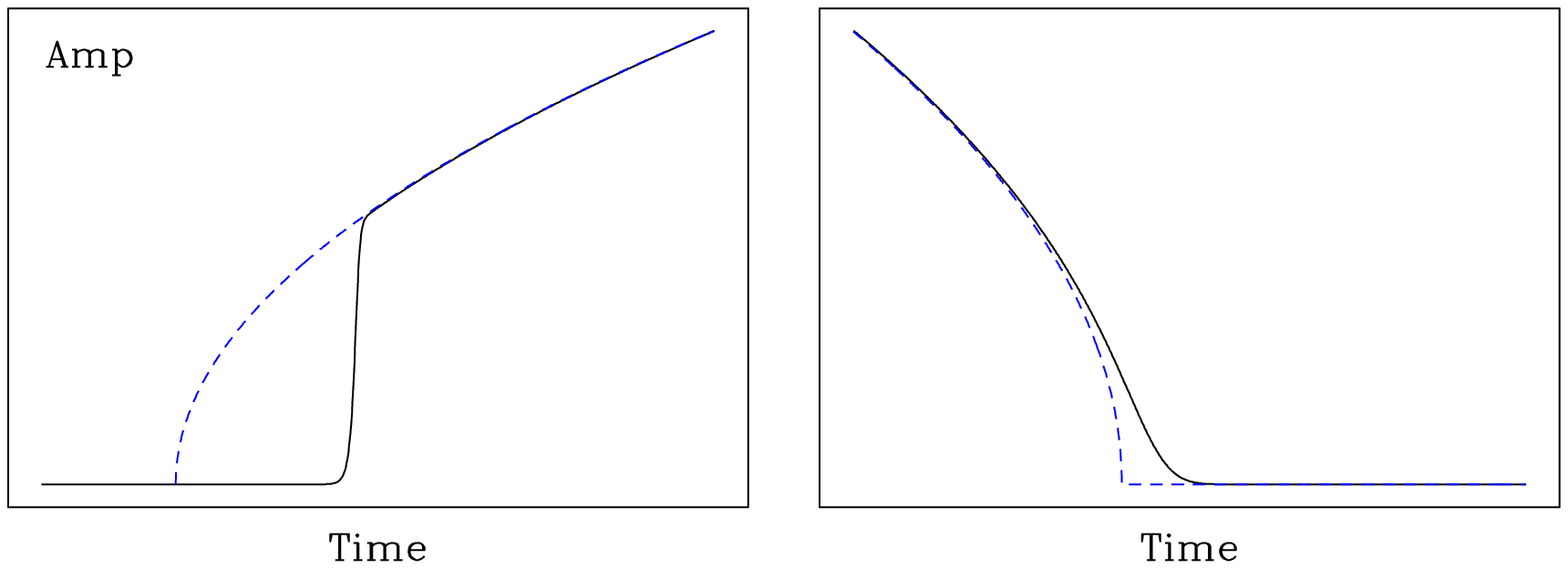}}
\end{center}
\vspace{0pt}
\noindent{\small Fig.~5: 
Solid line: behavior of the amplitude in the vicinity of the IS.
On the left, for a star entering the IS, and on the right, for a star 
leaving the IS.  
Dashed line: the limit cycle amplitude $A_{LC}$  that would be achieved
if $\tau_{evol}\gg\tau_{puls}$ everywhere.  }
\vspace{-2mm}
\end{figure}
First we have ascertained that they are not contaminated by nearby objects.
But that still leaves the possibility of binarity or rotation of a star with
spots.  Using bolometric corrections and \Teff\ from Bessell \& Germany (1999),
a distance modulus 18.55 and $A_v$=\th 0.18 (Keller \& Wood 2002; Marconi \&
Clementini 2005), we find for star 77.6940.14, $L$=2510\th\Lo, $M\th$=5\th\Mo,
\Teff =6200\th K and $ R\th$= 44\Ro; and for star 77.7548.11: $L\th$
=6730\th\Lo, $M\th$= 6.5\th\Mo, \Teff\th = 5400\th K, $R\th$=94\th\Ro.  Given
the observed periods $P$=1.32\th d and $P\th$= 6.03\th d, respectively, the
corresponding orbital radius of any (light) companion would be
$R_{orb}$=9\th\Ro\ and $R_{orb}\th$= 26\th\Ro, or 0.14 and 0.28 times the
stellar radius.  Under the above assumptions, the small orbital radii required
definitively rule out the binary hypothesis.  Similarly, rotation can also be
eliminated as an explanation because the rotation velocity derived from the
observed period and stellar radius is greater than the rotational breakup
velocity (by factors of 11.4 and 6.8, respectively).  We note, however, that
the above arguments fail if these two stars are foreground stars of much lower
luminosity than assumed.  Unfortunately, these stars are in a region of the LMC
HR-diagram where foreground contamination is large (the number of LMC and
foreground stars are comparable - see Alcock \etal 2000).  Radial velocity
measurements would be required to confirm LMC membership.  However, the fact
that they fall this close to the Cepheid PL relation suggests that they are 
probably LMC members.

Assuming these two stars are in the LMC, pulsation in strange modes is the
likely explanation for their variability.  The periods of the two stars are
indeed a factor of 5--6 lower than the periods of F Cepheid periods of equal
apparent magnitude, and in agreement with the expected periods of Strange
Cepheids (Buchler \& Kollath, 2001).  The amplitudes, $\sim$0.03\th mag, of
these stars are perhaps a little larger than the theoretical estimates, but
they are within their uncertainty.  A search for Strange Cepheids in the
remaining MACHO LMC fields would be desirable as it would strengthen the status
of this novel type of Cepheids.


One of us (JRB) gratefully acknowledges the hospitality of Mount Stromlo
Observatory.  We wish to thank Zolt\'an Koll\'ath for providing us with the
MUFRAN code.  This work has been supported by NSF (AST-0307281, OISE-0417772)
at the University of Florida.
This paper utilizes public domain data obtained by the MACHO Project, jointly
funded by the US DoE through the LLNL at the University of California
(W-7405-Eng-48), by the NSF through the UC Center for Particle Astrophysics
(AST-8809616), and by the Mount Stromlo and Siding Spring Observatory, part of
the ANU.  Support for OGLE was provided by Polish MNII (2P03D02124), NSF
(AST-0204908) and NASA (NAG5-12212).


{}

\end{document}